\documentclass[10pt,prd,aps,amsfonts,showpacs,longbibliography,twocolumn,notitlepage,superscriptaddress,nofootinbib]{revtex4-1}
\usepackage[dvipsnames]{xcolor}
\usepackage{multirow}
\usepackage{tabularx}
\usepackage{amsmath}
\usepackage{graphicx}
\usepackage{comment}
\usepackage{here}

\usepackage[colorlinks=true, urlcolor=violet, linkcolor=blue, citecolor=red, hyperindex=true, linktocpage=true]{hyperref}

\usepackage{slashed}

\allowdisplaybreaks

\numberwithin{thm}{section}

\newcommand{\tdif}[2]{\frac{d#1}{d#2}}
\newcommand{\pdif}[2]{\frac{\partial #1}{\partial #2}}

\newcommand{\mbb}{\mathbb}

\renewcommand{\thesection}{\arabic{section}}

\makeatletter
\renewcommand{\p@subsection}{}
\renewcommand{\p@subsubsection}{}
\makeatother

\usepackage{xcolor}
\usepackage{mathtools}
\usepackage{ragged2e}

\makeatletter 
    
\renewcommand\onecolumngrid{
\do@columngrid{one}{\@ne}%
\def\set@footnotewidth{\onecolumngrid}
\def\footnoterule{\kern-6pt\hrule width 1.5in\kern6pt}%
}

\renewcommand\twocolumngrid{
        \def\footnoterule{
        \dimen@\skip\footins\divide\dimen@\thr@@
        \kern-\dimen@\hrule width.5in\kern\dimen@}
        \do@columngrid{mlt}{\tw@}
}%

\makeatother   

\usepackage{dsfont}

\begin{document}

\title{Exact partition function of arithmetic Ising model}

\author{Anu DHOCHAK}
\author{Ken KIKUCHI}
\author{Shrinit SINGH}
\affiliation{International Center for Theoretical Sciences, Tata Institute of Fundamental Research, Shivakote, Hesaraghatta Hobli, Bengaluru North 560089, India}

\date{\today}

\begin{abstract}
We present a compact formula for the exact partition function of the $d$-dimensional arithmetic Ising model (AIM). For a $2\times2$ system, we express it analytically using the $q$-Hurwitz-Lerch zeta function and derive explicit forms for the free energy and entropy. Additionally, we find that the entropy increases at high temperatures, supporting the presence of entropic order.
\end{abstract}

\maketitle

\emph{Introduction.---}
Physical processes minimize free energy
\begin{equation}
    F=E-TS,\label{F}
\end{equation}
where $E$ is the energy, $T$ temperature, and $S$ entropy. For instance, matters realize phases with minimal free energy. Usual materials (such as water) form solid at low temperature and liquid at high temperature. At low temperature, the first term dominates. Thus, the molecules of water minimize kinetic energy by locking their positions sacrificing entropy. At high temperature, the second term dominates. Thus, the molecules of water maximize entropy by taking random configurations at the cost of kinetic energy.

The solid phase spontaneously breaks symmetries such as translation and rotation. They have specific shapes or orientations. Thus, the phase is called \textit{ordered}. On the other hand, the liquid phase has no preferred shapes or directions. The phase is called \textit{disordered}. The observation above suggests that matters are ordered at low temperature and disordered at high temperature.

However, some counterintuitive materials have also been known. For example, Rochelle salt \cite{K04} and the Pomeranchuk effect in $^3$He \cite{P50}. In these materials, one finds disordered phases at a low temperature and ordered phases at a higher temperature. The ordered phase is called the \textit{entropic order} \cite{HHKLP25}. Some minimal models including the arithmetic Ising model were constructed in \cite{HKLPS25}. (The ordered phases at arbitrarily high temperature were known in some quantum field theories. For example, see \cite{CCCKRS20,CCCKRS20',LRZ22,CRSS20,CCR20,BLS20,N21,CR21,AN21,HRS24,KP24}.) Furthermore, \cite{ANRS26} proved an existence of entropic order in generalized Ising models.

The entropic orders are believed to minimize free energy with enhanced entropy in the ordered phase. However, it is still unknown how the free energy and entropy change as a function of temperature in these models. In order to understand the entropic order, we need to verify enhanced entropy in the ordered phases. Therefore, in one of the minimal models known to realize entropic order, we compute its partition function, free energy, and entropy exactly. Our method is different from Onsager's tranfer matrix method \cite{O44,KO49}.\\

\emph{Computation.---} Let $L$ be a positive integer (which can be either even or odd). The arithmetic Ising model is defined on the $L\times L$ square lattice with the Hamiltonian
\begin{equation}
    H(\mathbf n,U,\mu):=U\sum_{x\sim y}n_xn_y+\mu\sum_xn_x,\label{Hamiltonian}
\end{equation}
where $x\sim y$ denotes the lattice sites $x,y$ are next-neighbors. $n_x$ are natural numbers (including zero), $n_x\in\mbb N$. The variables are subject to periodic boundary conditions in both directions. We also label the natural numbers with their positions in horizontal and vertical directions, $n_x=n_{i,j}$ with $i,j=1,2,\dots,L$. The periodic boundary conditions require $n_{i+L,j}=n_{i,j}=n_{i,j+L}$. This makes manifest a faithful symmetry $\mbb Z/L\mbb Z\times\mbb Z/L\mbb Z$
\begin{align*}
    \mbb Z/L\mbb Z:n_{i,j}&\mapsto n_{i+1,j},\\
    \mbb Z/L\mbb Z:n_{i,j}&\mapsto n_{i,j+1}
\end{align*}
of the system. The parameters $U,\mu$ are non-negative. For later convenience, we write the first and the second sums in the Hamiltonian $S_1,S_2$, respectively. The partition function is defined as its sum
\begin{equation}
\begin{split}
    Z(\beta,U,\mu):=e^{-\beta F(\beta,U,\mu)}&=\sum_{\{n_x\}} e^{-\beta H(\mathbf n,U,\mu)}\\
    &=\sum_{\{n_x\}}e^{-\beta US_1(\mathbf n)-\beta\mu S_2(\mathbf n)}.\label{partitionfunction}
\end{split}
\end{equation}
The partition function is bounded from above \cite{HKLPS25}:
\begin{equation}
    Z(\beta,U,\mu)\le Z(\beta,0,\mu)=\left(\frac1{1-e^{-\beta\mu}}\right)^{L^2}.\label{upperbound}
\end{equation}
The upper bound shows absolute convergence of the $L^2$ infinite sums because all summands of the partition function $e^{-\beta H(\mathbf n,U,\mu)}$ are positive.

Note that the parameters $U,\mu$ in the Hamiltonian appear in the product with $\beta$. This structure lets us remove one parameter. Explicitly, if we define
\[ \widetilde\beta:=\beta U,\quad\widetilde\mu:=\frac\mu U, \]
the partition function is rewritten as
\[ Z(\beta,U,\mu)=\sum_{\{n_x\}}e^{-\widetilde\beta S_1(\mathbf n)-\widetilde\beta\widetilde\mu S_2(\mathbf n)}=Z(\widetilde\beta,1,\widetilde\mu). \]
Since we always take $U>0$, the redefinition of parameters are well-defined. With this understanding, without loss of generality, we can set $U=1$. We work in this parametrization below. Accordingly, we write the partition function $Z(\beta,\mu)$, and similarly for the other functions.

Our primary goal of the paper is to compute the partition function (\ref{partitionfunction}) exactly. Our main result is the\\

\textbf{Theorem.} \textit{Let $L$ be a positive integer, $\beta,\mu>0$, and $z_{i,j}$ with $i,j=1,2,\dots,L$ (mod $L$) be formal variables. The partition function is given by}

\begin{align}
    Z(\beta,\mu) =\exp\left(-\frac\beta2\sum_{i,j=1}^L\pdif{}{\ln z_{i,j}}\left(\pdif{}{\ln z_{i-1,j}}+\pdif{}{\ln z_{i+1,j}}\right.\right.\nonumber\\
    \left.\left.+\pdif{}{\ln z_{i,j-1}}+\pdif{}{\ln z_{i,j+1}}\right)\right)\left.\prod_{i,j=1}^L\left(\frac1{1-z_{i,j}e^{-\beta\mu}}\right)\right|_{z_{i,j}=1}.\label{exactZ}
\end{align}\\

\textbf{Remark.} We generalize the theorem to $d$-dimensional arithmetic Ising model after the proof.\\

The theorem is a combination of propositions. First, we rewrite the problem. A straightforward way to compute the partition function $Z(\beta,\mu)$ is to perform the summation: $Z(\beta,\mu)=\cdots+e^{-\beta H(\mathbf n)}+\cdots$, where $\mathbf n=(n_{1,1},n_{1,2},\dots,n_{i,j},\dots,n_{L,L})$ is one configuration. Another way to perform the summation is to first collect configurations with the same energy. Denote the number of configurations with energy $E$ by $n_E$. Then, sum over all possible energies, $\sum_En_Ee^{-\beta E}$. Since the addition is associative, commutative, and absolutely convergent, the two methods provide the same result, i.e., $Z(\beta,\mu)=\sum_En_Ee^{-\beta E}$. A systematic way to organize configurations with the same energy is provided by the\\

\textbf{Proposition.} \textit{Let $\mu$ be irrational. $H(\mathbf{n_1},\mu) = H(\mathbf{n_2},\mu)$ implies $S_2(\mathbf{n_1}) = S_2(\mathbf{n_2})$. In other words, the matching of $S_2$ is a necessary condition for two configurations to have the same energy.}\\

\textbf{Proof.} Assume, for the sake of contradiction, that there exist two configurations $\mathbf n_1,\mathbf n_2$ such that $H(\mathbf n_1,\mu)=H(\mathbf n_2,\mu)$ and $S_2(\mathbf n_1)\neq S_2(\mathbf n_2)$. By definition, we have
\[ S_1(\mathbf n_1)+\mu S_2(\mathbf n_1)=S_1(\mathbf n_2)+\mu S_2(\mathbf n_2). \]
Since $S_2(\mathbf n_1)\neq S_2(\mathbf n_2)$, we can solve this for $\mu$
\[ \mu=\frac{S_1(\mathbf n_2)-S_1(\mathbf n_1)}{S_2(\mathbf n_1)-S_2(\mathbf n_2)}. \]
This contradicts our assumption that $\mu$ is irrational because both denominator and numerator are integers. $\square$\\

In order to employ the proposition, for a moment we assume $\mu$ is irrational. (We discuss how to relax the assumption after the computation.) Since configurations with the same energy have the same $S_2$, we can group them by setting $S_2$ to be a specific $N\in\mathbb N$. There are $\begin{pmatrix}L^2+N-1\\N\end{pmatrix}$ such configurations. Note that the configurations in general have different $S_1=m$. Let us denote the number of configurations with $S_1=m$ as $n_m$. Then, the partition function can be written as
\begin{equation}
\begin{split}
    Z(\beta,\mu)&=\sum_{N=0}^\infty e^{-\beta\mu N}\Big(n_0+n_1e^{-\beta\cdot1}+\cdots+n_{m}e^{-\beta m}\\
    &~~~~~~~~~~~~~~~~~~~~~~~~+\cdots+n_Me^{-\beta M}\Big).\label{partition2}
\end{split}
\end{equation}
By construction, we have $\sum_{m=0}^Mn_m=\begin{pmatrix}L^2+N-1\\N\end{pmatrix}$. The existence of the maximum $M$ of $S_1$ follows from the fact that the number is finite. (One can write the maximum $M$ in terms of $N$, but it turns out that it is not necessary.) Note that $L^2$ summations over natural numbers have reduced to just one summation.

How do we count the number $n_m$ of configurations with $S_1=m$? A useful observation is the\\

\textbf{Proposition.} \textit{Let $z_{i,j}$ with $i,j=1,2,\dots,L$ (mod $L$) be formal variables. There is a one-to-one correspondence between configurations $\mathbf n=(n_{1,1},n_{1,2},\dots,n_{i,j},\dots,n_{L,L})$ and monomials $z_{1,1}^{n_{1,1}}z_{1,2}^{n_{1,2}}\cdots z_{i,j}^{n_{i,j}}\cdots z_{L,L}^{n_{L,L}}$
}.\\

\textbf{Proof.} Given a configuration $\mathbf n=(n_{1,1},n_{1,2},\dots,n_{i,j},\dots,n_{L,L})$, we associate the monomial $z_{1,1}^{n_{1,1}}z_{1,2}^{n_{1,2}}\cdots z_{i,j}^{n_{i,j}}\cdots z_{L,L}^{n_{L,L}}$ and vice-versa. $\square$\\

Given a monomial $z_{1,1}^{a_{1,1}}\cdots z_{i,j}^{a_{i,j}}\cdots z_{L,L}^{a_{L,L}}$, one can read off $S_1$ from its powers. For example, the configuration corresponding to the monomial contributes $S_1=\sum_{i,j=1}^La_{i,j}(a_{i-1,j}+a_{i+1,j}+a_{i,j-1}+a_{i,j+1})$. This observation leads us to the\\

\textbf{Proposition.} \textit{Let $z_{i,j}$ with $i,j=1,2,\dots,L$ (mod $L$) be formal variables, and $\sum z_{1,1}^{a_{1,1}}z_{1,2}^{a_{1,2}}\cdots z_{L,L}^{a_{L,L}}$ be the sum of all monomials with degree $N$. Then, the derivative operator
\[
\begin{split}
&\exp\left(-\frac\beta2\sum_{i,j=1}^L\pdif{}{\ln z_{i,j}}\left(\pdif{}{\ln z_{i-1,j}}+\pdif{}{\ln z_{i+1,j}}\right.\right.\\
&~~~~~~~~~~~~\left.\left.\left.+\pdif{}{\ln z_{i,j-1}}+\pdif{}{\ln z_{i,j+1}}\right)\right)\right|_{z_{i,j}=1}
\end{split} \]
acting on $\sum z_{1,1}^{a_{1,1}}z_{1,2}^{a_{1,2}}\cdots z_{L,L}^{a_{L,L}}$ produces the bracket in (\ref{partition2}).}\\

\textbf{Proof.} The derivative operator is defined by
\[\begin{split}
\exp():=&\sum_{l=0}^\infty\frac{(-\beta/2)^l}{l!}\left[\sum_{i,j=1}^L\pdif{}{\ln z_{i,j}}\left(\pdif{}{\ln z_{i-1,j}}+\pdif{}{\ln z_{i+1,j}}\right.\right.\\
&\left.\left.+\pdif{}{\ln z_{i,j-1}}+\pdif{}{\ln z_{i,j+1}}\right)\right]^l.
\end{split} \]
Since it is a derivative operator, it acts on polynomials linearly. On a monomial $z_{1,1}^{a_{1,1}}\cdots z_{i,j}^{a_{i,j}}\cdots z_{L,L}^{a_{L,L}}$, it acts as
\begin{align*}
    &\exp()z_{1,1}^{a_{1,1}}\cdots z_{i,j}^{a_{i,j}}\cdots z_{L,L}^{a_{L,L}}\Big|_{z_{i,j}=1}\\
    =&\sum_{l=0}^\infty\frac{(-\beta/2)^l}{l!}\left[\sum_{i,j=1}^L\pdif{}{\ln z_{i,j}}\left(\pdif{}{\ln z_{i-1,j}}+\pdif{}{\ln z_{i+1,j}}\right.\right.\\
    &\left.\left.\left.+\pdif{}{\ln z_{i,j-1}}+\pdif{}{\ln z_{i,j+1}}\right)\right]^lz_{1,1}^{a_{1,1}}\cdots z_{i,j}^{a_{i,j}}\cdots z_{L,L}^{a_{L,L}}\right|_{z_{i,j}=1}\\
    =&\sum_{l=0}^\infty\frac{(-\beta/2)^l}{l!}\left[\sum_{i,j=1}^La_{i,j}\left(a_{i-1,j}+a_{i+1,j}+a_{i,j-1}+a_{i,j+1}\right)\right]^l\\
    \equiv&\exp\left(-\frac\beta2\sum_{i,j=1}^La_{i,j}\left(a_{i-1,j}+a_{i+1,j}+a_{i,j-1}+a_{i,j+1}\right)\right)\\
    &=e^{-\beta S_1(\mathbf a)}.
\end{align*}
Therefore, on the polynomial $\sum z_{1,1}^{a_{1,1}}z_{1,2}^{a_{1,2}}\cdots z_{L,L}^{a_{L,L}}$, we get the bracket. $\square$\\

Since the derivative operator is independent of $N$, it can be pulled out of the sum $\sum_N$. The partition function reduces to
\[ Z(\beta,\mu)=\exp()\sum_{N=0}^\infty e^{-\beta\mu N}\sum z_{1,1}^{a_{1,1}}z_{1,2}^{a_{1,2}}\cdots z_{L,L}^{a_{L,L}}\Big|_{z_{i,j}=1}. \]
The sum of all monomials with degree $N$ is given by the\\

\textbf{Proposition.} \textit{Let $t$ be a formal variable. The sum of all monomials $\sum z_{1,1}^{a_{1,1}}z_{1,2}^{a_{1,2}}\cdots z_{L,L}^{a_{L,L}}$ with degree $N$ is given by
\[ \sum z_{1,1}^{a_{1,1}}z_{1,2}^{a_{1,2}}\cdots z_{L,L}^{a_{L,L}}=\frac1{N!}\left.\tdif{^N}{t^N}\prod_{i,j=1}^L\left(\frac1{1-z_{i,j}t}\right)\right|_{t=0}. \]}\\ 

\textbf{Proof.} For each pair $(i, j)$, we expand the terms as a formal geometric series:$$\frac{1}{1 - z_{i,j}t} = \sum_{m=0}^\infty (z_{i,j}t)^m.$$
Taking the product over all $1 \le i, j \le L$, we obtain:
\[\begin{split}
    &\prod_{i,j=1}^L \frac{1}{1 - z_{i,j}t} = \prod_{i,j=1}^L \left( \sum_{m=0}^\infty z_{i,j}^m t^m \right) \\ &= \sum_{a_{1,1}, \dots, a_{L,L} \ge 0} \left( \prod_{i,j=1}^L z_{i,j}^{a_{i,j}} \right) t^{\sum_{i,j=1}^L a_{i,j}}.
    \end{split} \]
    By collecting terms of the same total degree in $t$, we can rewrite this product as a generating function:$$\prod_{i,j=1}^L \frac{1}{1 - z_{i,j}t} = \sum_{n=0}^\infty f_n t^n,$$where $f_n = \sum_{\sum a_{i,j} = n} \prod_{i,j=1}^L z_{i,j}^{a_{i,j}}$ is precisely the sum of all monomials in $z_{i,j}$ of total degree $n$. To extract the coefficient $f_N$, we apply the differential operator $\frac{1}{N!}\frac{d^N}{dt^N}$ to the series:
    \[ \begin{split} &\frac{1}{N!} \frac{d^N}{dt^N} \sum_{n=0}^\infty f_n t^n= \frac{1}{N!} \sum_{n=N}^\infty f_n \frac{n!}{(n-N)!} t^{n-N} \\ & = \sum_{n=N}^\infty \binom{n}{N} f_n t^{n-N}.\end{split}\]
    Evaluating this expression at $t = 0$, all terms with $n > N$ vanish due to the remaining powers of $t$. The only surviving term is for $n = N$:$$\left. \sum_{n=N}^\infty \binom{n}{N} f_n t^{n-N} \right|_{t=0} = \binom{N}{N} f_N = f_N.$$ Thus, the identity holds. $\square$\\

Combining the results so far, we arrive
\[Z(\beta,\mu)=\exp()\left.\left.\sum_{N=0}^\infty e^{-\beta\mu N}\frac1{N!}\tdif{^N}{t^N}\prod_{i,j=1}^L\left(\frac1{1-z_{i,j}t}\right)\right|_{t=0}\right|_{z_{i,j}=1}. \]
Note that the derivative with respect to $t$ and the summation over $N$ is nothing but the Taylor expansion. Since $e^{-\beta\mu}<1$ for positive $\beta\mu$, we get
\[ \sum_{N=0}^\infty e^{-\beta\mu N}\sum z_{1,1}^{a_{1,1}}z_{1,2}^{a_{1,2}}\cdots z_{L,L}^{a_{L,L}}=\prod_{i,j=1}^L\left(\frac1{1-z_{i,j}e^{-\beta\mu}}\right). \]
Plugging this into the expression above, we obtain (\ref{exactZ}).

Having obtained the expression, we can relax the assumption on $\mu$:\\

\textbf{Proposition.} \textit{The analytic expression $f(\beta, \mu)$ given in (\ref{exactZ}) is equal to the exact partition function $Z(\beta, \mu)$ for all $\beta > 0$ and $\mu > 0$.}\\

\textbf{Proof.} Fix $\beta > 0$. We have already established that the partition function $Z(\beta, \mu)$ agrees with the expression $f(\beta, \mu)$ for all positive irrational $\mu$. Define their difference as $g(\mu) = Z(\beta, \mu) - f(\beta, \mu)$. Both the analytic expression $f(\beta, \mu)$ and the formal partition function $Z(\beta, \mu) = \sum_{\{n_x\}} e^{-\beta H(\mathbf n,\mu)}$ are continuous with respect to $\mu$. Consequently, $g(\mu)$ is also continuous in $\mu$. Because $g(\mu) = 0$ on the positive irrationals, and the positive irrationals are dense in $(0, \infty)$, continuity dictates that $g(\mu) = 0$ for all $\mu > 0$. Thus, $Z(\beta, \mu) = f(\beta, \mu)$ for all $\mu > 0$. $\square$\\

This proves our theorem. We remark that the dimension entered only through the number of sites and the number of next-neighbors. None of the propositions leading to the theorem depend on the dimensions. Therefore, the theorem can be immediately generalized to the $d$-dimensional arithmetic Ising model:
\begin{equation}
\begin{split}
    Z(\beta,\mu)\\= &\exp\left(-\frac\beta2\sum_{i_1,i_2,\dots,i_d=1}^L\pdif{}{\ln z_{i_1,i_2,\dots,i_d}}\left(\pdif{}{\ln z_{i_1\pm1,i_2,\dots,i_d}}\right.\right.\\
    &~\left.\left.+\cdots+\pdif{}{\ln z_{i_1,i_2,\dots,i_{d-1},i_d\pm1}}\right)\right)\\
    &~~\times\prod_{i_1,i_2,\dots,i_d=1}^L\left(\frac1{1-z_{i_1,i_2,\dots,i_d}e^{-\beta\mu}}\right)\Bigg|_{z_{i_1,i_2,\dots,i_d}=1}.
\end{split}\label{exactZddim}
\end{equation}

Note that the derivative operator is originating from the $U$ term in the Hamiltonian. If we temporarily reintroduce the parameter $U$, we replace $\beta$ by $\beta U$. In the limit $U\to0$, the derivative operator becomes $\exp()=1$. Thus, we can set $z_{i,j}=1$ in the product. This reproduces the upper bound (\ref{upperbound}). This observation also provides an interpretation of our result. Without the $U$ term, increasing $n_{i,j}$ by one costs the chemical potential $\mu$ to the energy, and lowers the probability to realize the configuration by $e^{-\beta\mu}$. Since all lattice sites are independent without the interaction, we can sum over all possibilities by first fixing all sites but $(i,j)$, summing over $\sum_{n_{i,j}=0}^\infty e^{-\beta\mu n_{i,j}}$ to get a factor $1/(1-e^{-\beta\mu})$, and multiplying them to get the upper bound.

To compute physical quantities, we must evaluate the derivatives. While the results so far do not explicitly depend on the dimension $d$ or $L$, the derivatives introduce a dependence on them. We first explain the method for general case, and later demonstrate it in the special case $d=2=L$.\\

\textbf{Method.} First, we rewrite $w_{i,j}:=\ln z_{i,j}$. Next, we notice that the exponent can be viewed as a product $v^tAv$ of a vector $v:=(\partial_{1,1},\partial_{1,2},\dots,\partial_{L,L})^t$ -- ordered lexicographical as $(1,1),(1,2),\dots,(1,L),(2,1),\dots,(2,L),\dots,(L,L)$ -- where we use the abbreviation $\partial_{i,j}:=\pdif{}{w_{i,j}}$, and the adjacency matrix $A$. The matrix $A$ is symmetric in our notation to count each edge twice once for its one end and once for the other. Thus, it can be diagonalized by an orthogonal matrix $O$, $A=ODO^t$. (In the new basis $(k_{i,j})=O^t(w_{i,j})$, $w_{i,j}=0$ corresponds to $k_{i,j}=0$.) For $L\ge3$, the eigenvalues of the diagonal matrix $D$ are given by, for example, section 1.4.3 of \cite{BH12} ($L=1,2$ need separate treatment)
\[ \lambda_{k_1,k_2,\dots,k_d}=\sum_{i=1}^d2\cos\frac{2\pi k_i}L\quad(k_i=0,1,\dots,L-1,L\ge3). \]
Since $\beta\mu>0$ and we are only interested in the vicinity of $k_{i_1,\dots,i_d}=0$, we have $|e^{-\beta\mu+w_{i_1,\dots,i_d}(k)}|<1$. Thus, the resulting fractions can be written in terms of geometric series $\frac1{1-x}=\sum_{m\ge0}x^m$. Employing the linearity and $e^{a(\partial_x)^2}e^{bx}=e^{ab^2+bx}$, one can evaluate the exponentiated derivative operators.\\

The method works for $L\ge3$. For $L=1$, the computation is not difficult. For completeness, we demonstrate the method for $d=2=L$:\\

\textbf{Proposition.} \textit{For $d=2=L$, the partition function is given by}
\begin{equation}
    Z(\beta,\mu)=\left.\pdif{}z\left[\frac z{(1-q)^2}\Phi_q\left(z,2,\frac\mu2\right)\right]\right|_{z=e^{-\beta\mu},q=e^{-2\beta}},\label{analyticZdL2}
\end{equation}
\textit{where the $q$-Hurwitz-Lerch zeta function is defined as
\[ \Phi_q(z,s,a):=\sum_{n=0}^\infty\frac{z^n}{([n+a]_q)^s} \]
with $q$-number (for $q\neq1$)
\[ [n+a]_q:=\frac{1-q^{n+a}}{1-q}. \]}\\

\textbf{Proof.} The adjacency matrix is $A=2\begin{pmatrix}0&1&1&0\\1&0&0&1\\1&0&0&1\\0&1&1&0\end{pmatrix}$. (The overall coefficient 2 is understood as follows. For $L=2$, two edges, say, $(1,1)$-$(1,2)$ and $(1,2)$-$(1,1)$ are different, while they both contribute $n_{1,1}n_{1,2}$ to the Hamiltonian. Thus, the coefficient of $\partial_{1,1}\partial_{1,2}$ is $-2\beta$. With our normalization $-\beta/2$, we count the derivative operator four times. Distributing four symmetrically, we get the adjacency matrix with the overall coefficient.) It is diagonalized by $O=\frac12\begin{pmatrix}1&1&1&1\\-1&-1&1&1\\-1&1&-1&1\\1&-1&-1&1\end{pmatrix}$ to $D=\begin{pmatrix}-4&0&0&0\\0&0&0&0\\0&0&0&0\\0&0&0&4\end{pmatrix}$. In the new basis $\begin{pmatrix}k_{1,1}\\k_{1,0}\\k_{0,1}\\k_{0,0}\end{pmatrix}=O^t\begin{pmatrix}w_{1,1}\\w_{1,2}\\w_{2,1}\\w_{2,2}\end{pmatrix}$, the partition function reduces to
\[
\begin{split}
    Z(\beta,\mu)&=e^{2\beta\partial_{k_{1,1}}^2-2\beta\partial_{k_{0,0}}^2}\left(\frac1{1-e^{-\beta\mu+(k_{1,1}+k_{0,0})/2}}\right)^2\\
    &~~~~~~~~~~~~~~~~~~\left.\times\left(\frac1{1-e^{-\beta\mu+(-k_{1,1}+k_{0,0})/2}}\right)^2\right|_{k_{i,j}=0}.
\end{split}
\]
Note that there are no derivatives with respect to $k_{1,0},k_{0,1}$, and they can be set to zero. Since $|e^{-\beta\mu+(\pm k_{1,1}+k_{0,0})/2}|<1$, we can expand the fractions with $\frac1{(1-x)^2}=\sum_{m=0}^\infty(m+1)x^m$. We arrive
\begin{align}
    Z(\beta,\mu)&=e^{2\beta\partial_{k_{1,1}}^2-2\beta\partial_{k_{0,0}}^2}\nonumber\\
    &\times\sum_{m,n=0}^\infty(m+1)(n+1)\left.e^{-(m+n)\beta\mu+\frac{m-n}2k_{1,1}+\frac{m+n}2k_{0,0}}\right|_{k_{i,j}=0}\nonumber\\
    &=\sum_{m,n=0}^\infty(m+1)(n+1)e^{-(m+n)\beta\mu-2\beta mn}\nonumber\\
    &=\sum_{m=0}^\infty\frac{(m+1)e^{-\beta\mu m}}{(1-e^{-\beta\mu-2\beta m})^2}.\label{exactZsum}
\end{align}
It is real-analytic. By definition, the bracket in (\ref{analyticZdL2}) is given by
\[ []=\sum_{m=0}^\infty\frac{z^{m+1}}{(1-q^{m+\mu/2})^2}. \]
Setting $z=e^{-\beta\mu},q=e^{-2\beta}$ after the derivative, one obtains (\ref{exactZsum}). $\square$\\

\textbf{Remark.} $q$-deformed structures often arise in exactly solvable or integrable models. It would be interesting to explore hidden symmetries or integrability of the model.\\

\emph{Thermodynamic quantities.---}
With our exact analytic partition function (\ref{analyticZdL2}), the free energy and entropy are simply given by their definitions:
\begin{align*}
    F(\beta,\mu)&=-\frac1\beta\ln Z(\beta,\mu),\\
    S(\beta,\mu)&=-\pdif FT=\beta^2\pdif F\beta.
\end{align*}

In principle, these functions can be plotted, however, we are not aware of any software which supports a built-in $q$-Hurwitz-Lerch zeta function. Therefore, just for illustrative purposes, we calculate these functions numerically using the $\mathtt{NSum}$ function in Mathematica and plot them (we connect points for visibility). Because these numerical calculations are subject to errors originating from machine precisions, we use the graphs primarily to deduce the qualitative behaviors of the functions.
\begin{figure}[H]
    \centering
    \includegraphics[width=0.8\linewidth]{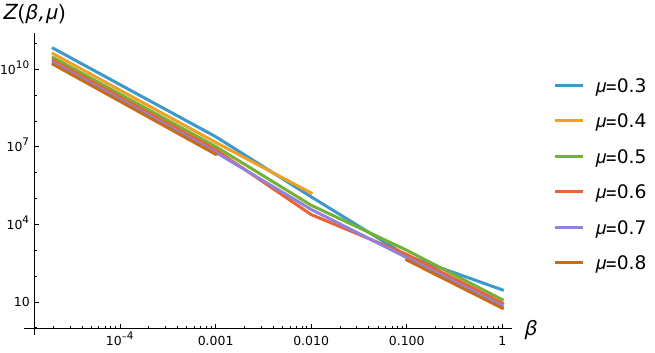}
    \caption{$Z(\beta,\mu)$ of $2\times2$ arithmetic Ising model}
    \label{Znum}
\end{figure}
\begin{figure}[H]
    \centering
    \includegraphics[width=0.8\linewidth]{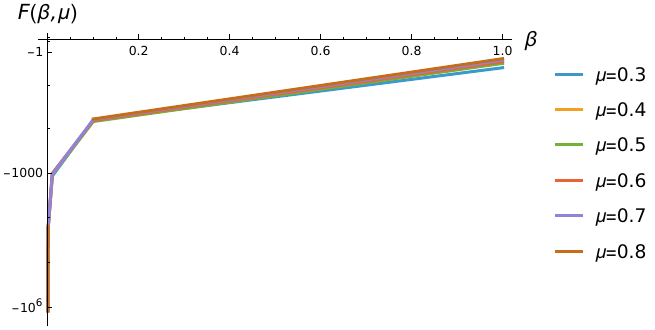}
    \caption{$F(\beta,\mu)$ of $2\times2$ arithmetic Ising model}
    \label{Fnum}
\end{figure}
\begin{figure}[H]
    \centering
    \includegraphics[width=0.8\linewidth]{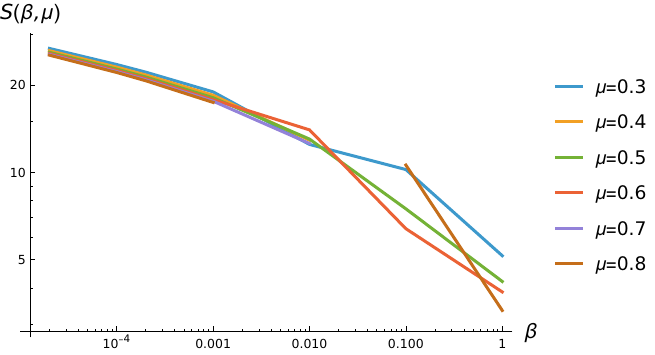}
    \caption{$S(\beta,\mu)$ of $2\times2$ arithmetic Ising model}
    \label{Snum}
\end{figure}
From the latter two graphs, one finds that, at high temperature (small $\beta$), the free energy is decreasing and the entropy is increasing. The enhanced entropy qualitatively supports the entropic order while there is no phase transition at finite $L$.

\section*{Acknowledgment}
We thank Enrico Andriolo and Zohar Komargodski for comments on a draft. We acknowledge support of the Department of Atomic Energy, Government of India, under project no. RTI4019.

\appendix
\setcounter{section}{0}
\renewcommand{\thesection}{\Alph{section}}
\setcounter{equation}{0}
\renewcommand{\theequation}{\Alph{section}.\arabic{equation}}

\end{document}